\documentclass[conference]{IEEEtran}
\makeatletter
\def\ps@headings{%
\def\@oddhead{\mbox{}\scriptsize\rightmark \hfil \thepage}%
\def\@evenhead{\scriptsize\thepage \hfil \leftmark\mbox{}}%
\def\@oddfoot{}%
\def\@evenfoot{}}
\makeatother
\usepackage{url}
\usepackage{color}
\usepackage{epsfig}
\usepackage{graphicx}
\usepackage{subfigure}
\usepackage{verbatim}
\usepackage{algorithm}
\usepackage{algorithmic}
\usepackage{array}
\usepackage{times}
 \usepackage{multirow}
\newcommand{\para}[1]{{\vspace{4pt} \bf \noindent #1 \hspace{10pt}}}

\newcommand{\htedit}[1]{{#1}}

\newenvironment{packed_itemize}{
\begin{itemize}
  \setlength{\itemsep}{1pt}
  \setlength{\parskip}{0pt}
  \setlength{\parsep}{0pt}
  \setlength{\headsep}{0pt}
  \setlength{\topskip}{0pt}
  \setlength{\topmargin}{0pt}
  \setlength{\topsep}{0pt}
  \setlength{\partopsep}{0pt}  
}{\end{itemize}}

\begin{document}


\title{Fast and Scalable Analysis of Massive Social Graphs}
\author{Xiaohan Zhao, Alessandra Sala, Haitao Zheng and Ben Y. Zhao \\
 	 Department of Computer Science, U. C. Santa Barbara, Santa Barbara, USA\\
 	\{xiaohanzhao, alessandra, htzheng, ravenben\}@cs.ucsb.edu}

\maketitle


\begin{abstract}
  Graph analysis is a critical component of applications such as online
  social networks, protein interactions in biological networks, and Internet
  traffic analysis.  The arrival of massive graphs with hundreds of millions
  of nodes, {\em e.g.} social graphs, presents a unique challenge to graph
  analysis applications.  Most of these applications rely on computing
  distances between node pairs, which for large graphs can take minutes to
  compute using traditional algorithms such as breadth-first-search (BFS).

  In this paper, we study ways to enable scalable graph processing on today's
  massive graphs.  We explore the design space of {\em graph coordinate
    systems}, a new approach that accurately approximates node distances in
  constant time by embedding graphs into coordinate spaces.  We show that a
  hyperbolic embedding produces relatively low distortion error, and propose
  {\em Rigel}, a hyperbolic graph coordinate system that lends itself to
  efficient parallelization across a compute cluster.  Rigel produces
  significantly more accurate results than prior systems, and is naturally
  parallelizable across compute clusters, allowing it to provide accurate
  results for graphs up to 43 million nodes.  Finally, we show that Rigel's
  functionality can be easily extended to locate (near-) shortest paths
  between node pairs.  After a one-time preprocessing cost, Rigel answers
  node-distance queries in 10's of microseconds, and also produces shortest
  path results up to 18 times faster than prior shortest-path systems with
  similar levels of accuracy.

\end{abstract}

\section{Introduction}
\label{sec:intro}

Fast and scalable analysis of massive graphs is a significant challenge
facing a number of application areas, including online social networks,
biological protein interaction networks, and analysis of the Internet router
backbone.  For example, a social game network might search
for ``central'' users to help deploy new games, while a social auction
site~\cite{overstock} wants to tell a buyer if a specific item is being
auctioned by someone in her social circles.  Ideally, such queries should be
answered quickly, regardless of the size of the graph, or even if graphs
themselves are changing over time.

Unfortunately, these goals are simply unattainable for today's massive
graphs.  This is because numerous graph analysis problems such as centrality
computation, node separation, and community detection all rely on the simple
{\em node distance} (length of shortest path) primitive, which scales badly
with graph size.  For graphs 
generated from social networks such as Facebook (500 million nodes), LinkedIn
(80 million) and Twitter (100 million), computing the shortest path distance between a
single pair of nodes can take a minute or more using traditional algorithms
such as breadth-first-search (BFS)~\cite{shortpath}.  Similarly, variants such as
Dijkstra and Floyd-Warshall also fail to scale to these graph sizes.

Without an efficient alternative for node distance computation, recent work
has focused on exploring efficient approximation
algorithms~\cite{shortpath,NSI06,orion10}.  Our prior work~\cite{orion10},
described the idea of {\em graph coordinate systems}, which embeds graph
nodes into points on a coordinate system.  The resulting coordinates can be
used to quickly approximate node distance queries on the original graph.
Our initial system, which we refer here to as Orion, was a centralized
system that approximated node distances by mapping nodes to the Euclidean
coordinate system.  It has several limitations in practice.  First, Orion's
initial graph embedding process is centralized and computationally expensive,
which presents a significant performance bottleneck for larger graphs.
Second, Orion's results produce error rates between 15\% and 20\%, which
limits the types of applications it can serve.  Finally, it is unable to
produce actual paths connecting node pairs, which is often necessary for a
number of graph applications.

In this work, we seek to extend work on {\em graph coordinate systems} by
developing a practical system that provides significant improvement in
accuracy, scalability, and extended functionality.  We systematically explore
decisions in the design of a graph coordinate system, and make two key
observations.  \emph{First}, we propose to extend our work on graph
coordinate systems, by embedding large graphs in a hyperbolic space for lower
distance distortion errors.  Our embedding algorithm naturally parallelizes
the costly embedding process across multiple servers, allowing our system to
quickly embed multi-million node graphs.  \emph{Second}, we propose a novel
way to use graph coordinates to efficiently locate shortest paths between
node pairs.  The result of our work is {\em Rigel}, a hyperbolic graph
coordinate system that supports queries for both node distance and shortest
paths on today's large social graphs.  After a one-time, easily
parallelizable, preprocessing phase, Rigel can resolve queries in tens of
microseconds, even for massive social graphs up to 43 million nodes.


Our paper describes four key contributions. 
\begin{packed_itemize}
\item In Sections~\ref{sec:hyper} and \ref{sec:para}, we describe the
  detailed design of Rigel, and show how we can minimize embedding time by
  effectively parallelizing the most computationally expensive parts of the
  graph embedding process.
\item We evaluate a distributed prototype of Rigel using social graphs of
  different sizes from several OSNs, including Facebook, Flickr, Orkut,
  LiveJournal, and Renren.  Our results show that Rigel achieves consistently
  improved accuracy compared to Orion, and scales to large graphs of up to 43
  million nodes.
\item In Section~\ref{sec:app}, we implement three different social graph analysis
  applications on top of the Rigel system.  Our results illustrate both the
  accuracy and scalability of the Rigel system for use in real graph analysis
  applications.
\item Finally, we propose an approach to approximate shortest paths for any
  node pair using graph coordinates.  We compare Rigel's shortest path
  results to those from recently proposed techniques.  Rigel paths provide
  accuracy similar to the most accurate of prior schemes, while resolving
  queries up to 18 times faster.
\end{packed_itemize}

\begin{table}[t]
\begin{small}
\centering
\begin{tabular}{| c | c c c|}
\hline
Graphs & Nodes & Edges & Avg. Path Len.\\
\hline
Egypt & 246K & 1,618K & 5.0\\
Norway & 293K & 5,589K & 4.2\\
L.A. & 275K & 2,115K & 5.2\\
\hline
Flickr & 1,715K& 15,555K & 5.1 \\
Orkut & 3,072K & 117,185K & 4.1\\
Livejournal & 5,189K & 48,942K & 5.4 \\
Renren & 43,197K & 1,040,429K & 5.0\\
\hline
\end{tabular}
\caption {A variety of social graphs used in our work.}
\label{tab:graphs}
\end{small}
\end{table}

\subsection{Social Network Graph Datasets}  
Throughout our paper, we use a number of anonymized social graph datasets
gathered from measurements of online social networks to guide and evaluate
our system design.  We utilize a total of 7 social graphs, ranging in size
from 246,000 nodes and 1.6 million edges, to 43.2 million nodes and 1 billion
edges.  We list their key characteristics in Table~\ref{tab:graphs}.

Three of these graphs, Egypt, Los Angeles (LA) and Norway, are Facebook
regional networks shared by the authors of~\cite{interaction}.  The remaining
four graphs are significantly larger graphs crawled from the Flickr, Orkut,
LiveJournal, and Renren social networks, each with millions of nodes and
edges.  We use them to test the efficiency and scalability of our system.
The Livejournal, Flickr and Orkut are datasets shared by the authors
of~\cite{mislove07}.  With 43 million nodes and more than 1 billion edges,
our largest dataset is a snapshot of Renren, the largest online social
network in China.  We obtained this graph after seeking permission from
Renren and the authors of~\cite{jiang10}.  While these graphs are still
significantly smaller than the current user populations of Facebook (600
million) and LinkedIn (80 million), we believe our graphs are large enough
to demonstrate the scalability of our mechanisms.

\section{Background and Related Work}
\label{sec:back}

Our goal is to develop a practical system that quickly answers node distance
queries for today's massive social graphs.  To do so, we will use our proposed
concept of {\em graph coordinate systems} (GCS), an approach that
tolerates an initial computational overhead in order to provide
node-distances approximations that take constant time regardless of graph
size.  In this section, we introduce the concept of graph coordinate systems,
and related work on graph embedding and social networks.


\subsection{Background}
\begin{figure}[t]
\centering
   \epsfig{file=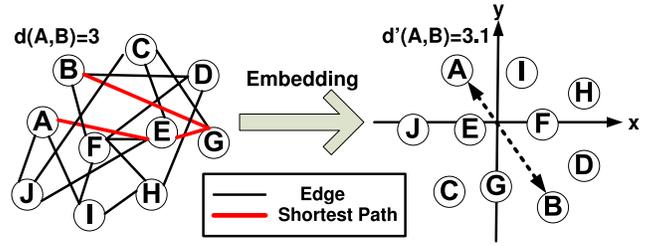,width=3in}
   \caption{An example of graph embedding to an Euclidean space. For example,
   the shortest path distance between nodes $A$ and $B$ is 3 in the graph
   (left), and the Euclidean distance between their coordinates is 3.1 (right).}
   \label{fig:embed}
\end{figure}

Graph coordinate systems, a concept first proposed in
Orion~\cite{orion10}, seek to provide accurate estimates of distances 
between any pair of graph nodes. At a high level, this approach captures the
complex structure of a high dimensional graph, and computes a lossy
representation for it in the form of a fixed position for each graph node in
a coordinate space.  Each node's coordinate position is chosen such that its
distance to another node in the coordinate space matches its real shortest
path distance to that node in the actual graph.  In Figure~\ref{fig:embed} for
example, the shortest path distance between nodes $A$ and $B$ is 3 in the
graph, and the Euclidean distance between their coordinate positions is 3.1.

\para{Pros and Cons.}  The advantage of using a GCS is that,
once a graph is embedded, the system can answer each node distance query
using a small amount of time independent of the graph size, {\em
  i.e.} $O(1)$ time.  In practice, each query takes only a few microseconds
($\mu s$) to compute.  This is very attractive for applications that require
large numbers of node distance computations, such as computing graph-wide
metrics like graph diameter and average path length.  To process queries on a
given graph $G$, however, a GCS must first
compute a one-time embedding of $G$ into the coordinate space, {\em i.e.}
compute the coordinate positions of each graph node.  This initial step can be
computationally expensive, and scales roughly linearly with graph size, {\em
  i.e.} $O(n)$ for a graph with $n$ nodes.  Finally, a graph coordinate
system provides good approximations to graph queries, but does not provide
perfect results. 

\para{Goals.}  We focus on two goals in our exploration of the GCS design
space.  First, we seek to optimize the graph embedding to 
maximize accuracy.  Second, since graph embedding is by far the biggest
source of computational cost in a GCS system, our goal is to ensure that we can take
advantage of distributed computing resources, {\em e.g.} server clusters, to
ensure scalability as graphs continue to grow.

\subsection{Work on Embedding in Geometric Spaces}
\label{sec:related}

Embedding techniques have been used in a variety of application contexts.  
The most recent and well-known use of embedding techniques was in the context
of network coordinate systems used to estimate Internet latencies without
performing exhaustive end-to-end measurements~\cite{GNP,IDMaps,Vivaldi}. In
contrast, we are interested in finding an efficient embedding metric that
preserves shortest paths on large complex graphs, such as those derived from
social networks.


We highlight three popular geometric spaces: Euclidean,
Spherical and Hyperbolic, and summarize prior experiences with these spaces
from both measurement and theoretical studies.

\para{Euclidean.}  Euclidean embedding was first used on simple graphs such
as planar graphs and trees~\cite{small_distortion}.  It was widely used to
predict ``distances'' such as routing latency between Internet
hosts~\cite{GNP,Vivaldi}.  For example, GNP~\cite{GNP} is a centralized system that uses
a small number of public landmarks to embed all Internet hosts in the space.
Similar systems proposed later include those using Lipschitz
embedding~\cite{TangCrovella:imc03}, a spring force model~\cite{Vivaldi} and
most recently a system using Euclidean Big-Bang
Simulation~\cite{BBS_Euclidean}.  These systems calibrate nodes' geometric
positions based on distances, {\em e.g.} Internet round-trip time (RTT),
which are measured in a distributed manner.  Still later work proposed bounds on the
distortion of Euclidean embedding.  To the best of our knowledge, J. R. Lee's
recent result~\cite{volume_distortion} proves the tightest upper bound,
$O(\sqrt{\log n}\log{\log n})$ for an $n$-point Euclidean embedding.

\para{Spherical.}  Vivaldi~\cite{Vivaldi_hyperbolic} was the first to
investigate the accuracy of embedding a network into a spherical coordinate
space.  While morphing on spherical spaces is widely used in computer
vision~\cite{Morphing_sherical}, there is little theoretical work
investigating spherical embedding.

\para{Hyperbolic.}  A hyperbolic space can be thought of a space with a
tightly connected core, where all paths between nodes pass through.
Intuitively, both social graphs and the Internet topology should fit this
model well, since they both feature highly connected graph cores.
Experimental systems for embedding Internet distances~\cite{Crowcroft_IMC05,
  BBS_Hyperbolic,Vivaldi_hyperbolic} generally showed improved accuracy over
analogous systems that used Euclidean spaces.

There is limited work on theoretical characteristics of Hyperbolic spaces
embedding.  In the context of ad hoc wireless networks, Kleinberg proved that
a {\em greedy hyperbolic embedding} yields routes with low
stretch~\cite{Geo_routing_hyperbolic}, where greedy embedding is a graph
embedding with the following property: for any pair of nodes $(u,v)$, there
is at least one neighbor of node $u$ closer to node $v$ than node $u$ itself.
A recent work~\cite{Hyp_Greedy_Infocom09} improves the greedy embedding
algorithm for dynamic graphs, and proposes a modified greedy
routing algorithm for message routing. 

While these projects are about Hyperbolic embedding algorithms, they either
focus on graphs in the context of routing in wireless networks or on small
synthetic graphs ($\sim$50 nodes as in~\cite{Hyp_Greedy_Infocom09}).  A
later project~\cite{greedyforward10} proposes a graph model using Hyperbolic
spaces that is capable of producing synthetic graphs with scale-free
structural properties.  Unlike our work, this project aims to generate
synthetic graphs instead of embedding real graphs.
\subsection{Social Network Applications and Studies}
Here we 
briefly summarize other related projects on social applications 
and social network measurements.


\para{Shortest-path based Applications.}
Recently, social networks have inspired a numerous security protocols and
social applications in a number of fields. In Section~\ref{sec:app}, we will
evaluate our proposed system using three of the most common social analysis
applications: graph separation metrics, graph centrality, and distance-ranked
social search~\cite{shortpath,socialsearch}.

There are many other social applications relying on shortest path
computations.  For instance, information dissemination~\cite{fast_max_influ}
can leverage distances between nodes to find the most influential
nodes. Community detection algorithms on social graphs (see a taxonomy
from~\cite{santo10}) can benefit from shortest path distances between nodes
to classify them in different clusters. Furthermore, algorithms for detecting
Sybil attacks are similar to community detection strategies~\cite{sybil10},
which make them suitable candidates to leverage our system.  Neighborhood
function~\cite{anf02} uses node distance distributions to predict whether two
graphs are similar or not. Mutual friends detection computes the mutual
friends between social users. Users in the Overstock social auction site
query the social graph to see how they are connected to sellers of a given
product~\cite{overstock}.  All these applications rely heavily on shortest path
computations, and therefore can benefit significantly from our system.

\para{Studies of Online Social Networks.}
Recently, a number of large measurement studies have studied the structure of
online social networks through graph measurement and analysis.  For example,
Mislove et al.  published a comprehensive paper to analyze data crawled from
Flickr, Livejournal, Orkut and Youtube~\cite{mislove07}. Wilson et
al. generated large social graphs and interaction graphs by crawling the
Facebook network~\cite{interaction}.  Jiang et al.~\cite{jiang10} used the
same methodology to generate a large social graph of 43 million users on
Renren, the Chinese Facebook clone.  Finally, Twitter was analyzed in
\cite{twitter_media}, and other studies modeled behavior of social network
users using network level data measurements~\cite{fabricio09,fabian09}.

\section{A Hyperbolic Graph Coordinate System}
\label{sec:hyper}
A number of recent projects have shown that hyperbolic spaces can more
accurately capture distances on a network
graph~\cite{BBS_Hyperbolic,Hyp_Greedy_Infocom09,greedyforward10}.  We also empirically
compute distortion metrics~\cite{distortion} on our social graphs for
different coordinate systems in Table~\ref{tab:indmet}, and find that the hyperbolic space is in fact significantly more
accurate than Euclidean and spherical alternatives

\begin{table}[t]
\centering
\begin{small}
\begin{tabular}{|c|c c c|c|}
\hline
\multirow{2}{*}{Metrics} & \multirow{2}{*}{Euclidean} &
\multirow{2}{*}{Hyperbolic} & \multirow{2}{*}{Spherical} & Ideal \\
&&&&Value \\
\hline
ARE & 0.16 & {\bf 0.10} & 0.36 &\multirow{2}{*}{0} \\
AAE & 0.78 & {\bf 0.50} & 1.83 & \\ \hline
AER & 0.97 & {\bf 1.00} & 0.91 &\multirow{4}{*}{1}  \\
ACR & 1.07 & {\bf 1.02} & 1.72&  \\
ASPD & 1.19 & {\bf 1.11} & 1.96 & \\
SD  & 58.46 & {\bf 30.63} & 134173.04 & \\
\hline
\end{tabular}
\caption {\small Evaluating different embedding spaces via several metrics on
  the Facebook LA graph. Note the following acronyms: average
  relative error (ARE),  average absolute error (AAE), average expansion
  ratio (AER), average contraction ratio (ACR), average symmetric pair
  distortion (ASPD), and space distortion (SD).}
\label{tab:indmet}
\end{small}
\end{table}

In this section, we describe {\em Rigel}, a hyperbolic graph coordinate system (GCS) for
estimating node distance queries.  Before answering queries on a particular
graph, the graph must first be embedded into a hyperbolic coordinate space, a
process that involves computing ideal coordinate values for each node in the
graph.  We describe hyperbolic coordinate computation in
Rigel, present details of Rigel's graph embedding process, and explore the
impact of system parameters on embedding accuracy.  Wherever possible, we
compare Rigel's results directly to comparable results obtained from running
Orion~\cite{orion10}, our prototype GCS based on Euclidean coordinates.

\subsection{Distance Computation in the Hyperboloid}
There are five known ``Hyperbolic models'' that have been proposed for
different purposes and graph structures, including the Half-plane, the
Poincar\'e disk model, the Jemisphere model, the Klein model and the
Hyperboloid model~\cite{BBS_Hyperbolic}.  Each model is a
different method of assigning coordinates and computing distances over the
same hyperbolic structure.  Since choosing a model fundamentally changes how
graphs can be embedded, it is currently unknown how the choice of
models affects embedding distortion.

In designing Rigel, we
chose the {\em Hyperboloid} model for two practical reasons.  First,
computing distances between two points in this model is
computationally much simpler than alternative models.  Second, the
computational complexity of calculating distances is independent of the space
curvature.  This gives us additional flexibility in tuning the structure of
the hyperbolic space for improved embedding accuracy. 

The curvature parameter $c$~($c\leq 0$ in our model) is another important parameter in the definition
of the Hyperbolic space~\cite{BBS_Hyperbolic}. When $c=0$, the Hyperbolic
space reduces to the Euclidean space. The choice of $c$ also has significant impact on
the level of distortion between the real node distances and their images on
the Hyperbolic space.  For a Hyperboloid model with curvature $c$, the
distance between two $n$-dimension points $x=(x_1,x_2,\ldots,x_n)$ and
$y=(y_1,y_2,\ldots,y_n)$ is defined as follows:

\begin{small}
\begin{equation}
\label{eq:dis}
\delta(x,y)=arccosh\left(\sqrt{(1+\sum_{i=1}^n{x_i^2})(1+\sum_{i=1}^n{y_i^2})}-\sum_{i=1}^n{x_iy_i}\right)
\cdot |c|
\end{equation}
\end{small}
As we will empirically show in Section~\ref{sec:curvature}, smaller absolute values of 
$c$~(when 
 $5 \geq |c|\geq 1$) produce lower distortion.

\subsection{Computing a Hyperbolic Embedding}
\label{sec:bootstrap}

We now describe a basic (centralized) algorithm for embedding a graph into
our Hyperbolic space.  At a high level, we follow the ``landmarks'' approach
proposed in~\cite{orion10}, where we first choose a small number of $l$
nodes as landmarks, where $l \ll N$ and $N$ is the number of nodes in $G$.
We first use a global optimization algorithm to fix the coordinates of these
landmarks, such that their distances to each other in the coordinate space are
as close as possible to their matching path distances in the
graph.  We refer to this step as ``bootstrapping.'' Once the
landmarks are set, we compute the positions of all remaining nodes, such
that each node's distances to all landmarks in the coordinate space closely
match its actual node distances to those landmarks in the graph.  

The rationale behind this approach is that computing ``ground truth,'' {\em
  i.e.} the shortest path length between any two nodes, is an expensive task.
This is unlike other embedding applications, {\em e.g.} Internet latencies,
where a single ``Ping'' would get the true distance between 2 nodes. Thus
``calibrating'' node positions in a pairwise fashion would generate a large
number ($O(N^2)$) of breadth-first-search (BFS) computations.  By choosing a
small, constant number of landmarks, we only need to compute a BFS tree for
each landmark.  The resulting values represent shortest path lengths from all
remaining nodes to these landmarks, and are sufficient to calibrate their
coordinate values.  As in~\cite{orion10}, we choose the landmarks as nodes
with highest degree, as a way to efficiently approximate nodes with high
centrality. 

Next, to compute the coordinate position for a graph node, we randomly 
select $16$ 
out of the $l$~($l=100$) landmarks.  Recall that we computed a global BFS from each
landmark to all nodes in the graph during the bootstrapping step.  Thus we
know the actual node distances in the graph between 
the new node and each of its $16$ selected landmarks.  We apply the Simplex
method~\cite{simplex} to compute an optimal coordinate such that distances
between the node and its landmarks in the coordinate space match the known
node distances.

\para{Optimizing Local Paths.}
It has been shown in Internet embedding systems~\cite{Crowcroft_IMC05} that
the largest errors are introduced when estimating paths or node distances for
nearby nodes, {\em i.e.} nodes separated only by 1 or 2 hops.  In the context
of graphs, this is an easy limitation to overcome, since 1-hop neighbors are
easily accessible via graph representations such as edge lists or adjacency
matrices.  Rigel uses local neighbor information to augment the node
knowledge about its close-by topology.  Before answering a query for a pair
of nodes, Rigel first checks their adjacency lists to detect if they are
direct neighbors or 2 hop neighbors (share a node in their adjacency list).

This additional memory access increases Rigel's per-query latency, but is
still a worthwhile tradeoff for two reasons.  First, accuracy in resolving
local graph queries is critical to many graph operations.  Second, we will
show later that even after the optimization, overall latency for each query is
still limited to tens of microseconds for our graphs.

\subsection{Embedding Accuracy on Real Graphs}
\label{sec:eval}
We now investigate the impact on embedding accuracy by two parameters,
curvature of the space \htedit{$c$} and number of dimensions of the space \htedit{$n$}.  We report
experimental results using three Facebook datasets presented in
Table~\ref{tab:graphs}.  The results on the remaining graphs are consistent
with these results, and are omitted for brevity.  Next, we take a closer look
at the magnitude of approximation errors as a function of the actual path
lengths, and find that as expected, relative errors are highest for node
pairs already close by in the graph.  


\subsubsection{Impact of Curvature and Dimension}
\label{sec:curvature}
In order to derive the parameters that maximize the accuracy of our system,
we evaluate the impact of two important parameters of Hyperbolic space:
curvature and number of dimensions.

\para{Impact of Curvature.} 
The curvature $c$ of a Hyperbolic space is an important parameter that
determines the structure of the space.  We build different Hyperbolic spaces
using curvature values that range from $-50$ to $0$, and investigate the
effect on the accuracy of the distance estimation using our three Facebook social graphs. 

\begin{figure}[t]
\centering
   \epsfig{file=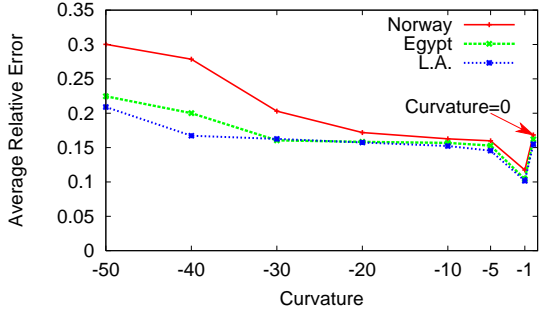,width=3in}
\vspace{-0.1in}
\caption{Impact of hyperbolic curvature on accuracy.}
\label{curve}
\end{figure}

Figure~\ref{curve} plots the average relative error when the curvature
varies between $-50$ and $0$.  When the curvature is $0$, the Hyperbolic
space is equivalent to an Euclidean space.  We include this value as the
rightmost point in our plot.  From our results, we see that the average error
decreases significantly as the curvature approaches $-1$.  We performed further
fine grain tests with curvature values around $-1$, and find that the
accuracy of our system reaches a plateau near $-1$.  Results at curvature of
-1 are $30$\% more accurate than results from an Euclidean system, shown in
the plot as curvature of $0$. Thus we use the curvature value at $-1$
in the rest of this paper.

\para{The impact of Dimensions.} The number of dimensions of a geometric
space plays an important role in determining the accuracy level in the
estimate of distances between nodes.  Therefore, we vary the number of
dimensions from $2$ to $14$ and evaluate the resulting accuracy.
Increasing dimensions reduces the error from more than $0.2$ to
about $0.1$, with most of the significant improvement occurring between $2$
and $6$ dimensions.  Since the results are not new, we omit the figure for
brevity.  Since the number of dimensions is a linear factor in the
computational complexity of the Simplex method used in our embedding, we need
to balance prediction accuracy against computational complexity.  We find a
sweet spot close to $10$ dimensions, where the accuracy has essentially
reached a plateau.  Thus we also use $10$-dimension for our Hyperbolic
system.  This has the added benefit of providing a fair and direct comparison
with our instance of Orion, which uses a 10-dimension Euclidean space.


\begin{figure}[t]
\centering
   \epsfig{file=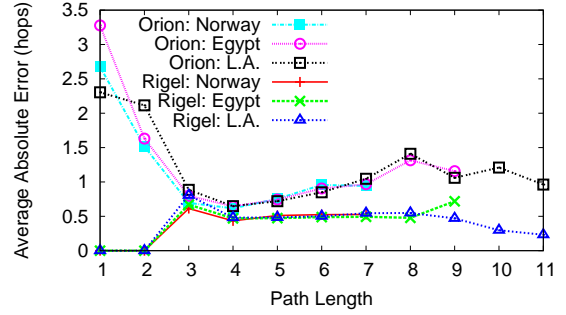,width=3in}
\vspace{-0.1in}
   \caption{Average absolute errors for paths of different lengths. 
   The top three lines are from Orion with errors in $[0.6,3.4]$. 
   The bottom three lines are from Rigel with errors in $[0,0.9]$.}
   \label{fig:newab}
\end{figure}

\begin{figure}[t]
\centering
   \epsfig{file=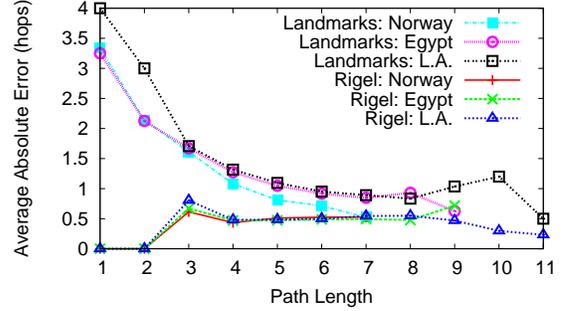,width=3in}
\vspace{-0.1in}
\caption{Average absolute errors comparing Rigel and the ``Landmark'' scheme
  from CIKM 2009.  The ``Landmark'' scheme shows errors in the range
  $[0.5,4]$, and Rigel in $[0,0.9]$.}
   \label{fig:24vsrigel}
\end{figure}

\begin{table}[t]
\centering
\begin{tabular} {|c|c c c|c|}
\hline
Graphs & Orion & Rigel-S & Rigel & BFS \\
\hline
Egypt & 0.2$\mu$s & 0.33$\mu$s & 6.8$\mu$s & 0.75s \\
L.A. & 0.18$\mu$s & 0.33$\mu$s & 8.5$\mu$s & 1.027s\\
Norway & 0.19$\mu$s & 0.33$\mu$s & 17.8$\mu$s & 1.44s\\
\hline
\end{tabular}
\caption{Response time for Orion, Rigel-S, Rigel and BFS.}
\label{tab:restime}
\end{table}

\subsubsection{Accuracy and Per-query Latency}
In this section, we examine accuracy as a function of path length, and
also compare per-query latency across a number of systems.  In all cases, we
use a 10-dimensional Hyperbolic space with curvature of $-1$.

\para{Accuracy vs Path Length.}  We explore the accuracy of predictions for
paths of different lengths.  Our accuracy breakdown tells us how our node
distance approximations perform, as a function of how far away the two nodes
are apart in the actual graph. 

Figure~\ref{fig:newab} shows the average absolute errors per path length on
three Facebook graphs by leveraging two embedding systems: Orion (using an
Euclidean space) and Rigel. The bottom three lines are the results of Rigel
where the average absolute error per path length ranges between 
$0$ and $0.9$. Comparing Rigel to Orion~(the top three lines in
Figure~\ref{fig:newab}), we confirm a noticeable improvement. Indeed, Orion
presents an average absolute error per path length between $0.6$ 
and $3.4$ which is significantly higher than Rigel. 
This shows that using hyperbolic spaces clearly has a significant impact on
accuracy. Also note that Orion produces extremely
large errors for close node pairs.  These errors are completely eliminated
by Rigel's local path optimization. 

We also compare Rigel's accuracy against the ``Landmark'' scheme proposed
in~\cite{shortpath}.  Our results in Figure~\ref{fig:24vsrigel} show that 
Rigel significantly outperforms~\cite{shortpath} regardless of the real
node distance between the nodes.  A comparison with Figure~\ref{fig:newab}
shows that Orion also provides slightly better accuracy than
\cite{shortpath}. 

\para{Query Latency.} Table~\ref{tab:restime} shows the average per-query
response time required to compute the distance of two randomly selected nodes
using Orion, Rigel, and BFS.  We also plot the query time of Rigel without
the local path optimization, and label it as ``Rigel-S.''  Rigel-S requires
slightly longer time than Orion, because of the increased complexity of the
hyperboloid coordinate computation.  Memory access in Rigel's local path
optimization adds several microseconds to each query.  But overall, Rigel's
per-query time is still \htedit{$5$} orders of magnitude faster than BFS.

\section{Embedding Massive Graphs}
\label{sec:para}
While we have described basic techniques to embed large graphs to a
hyperbolic space, preliminary evaluation of our system revealed a significant
challenge.  Because the complexity of initial embedding scales linearly with
the number of nodes in the graph, embedding a graph with multi-million ({\em
  e.g.} 43M) nodes can take up to a week to complete.  This processing
overhead presents a significant performance bottleneck, and the final
limitation that prevents the practical application of Rigel on today's
massive social graphs.

In this section, we describe a natural way to address this limitation by
leveraging the availability of distributed server clusters.  Rigel's
embedding process is easily parallelizable across multiple servers, allowing
us to reduce embedding time from a few days to a few hours using a cluster of
50 commodity servers.  We refer to this optimization as ``parallel Rigel.''
Here, we describe mechanisms involved in parallelizing Rigel's embedding
process, and then evaluate its impact using four large social graphs.

\begin{table*}[t]
\centering
\begin{tabular}{| c| c c | c | c c | c c |}
\hline
Graphs & \multicolumn{2}{|c|}{Bootstrap (hours)}  & Graph Partitioning
(hours)& \multicolumn{2}{|c|}{Embedding (hours)} & \multicolumn{2}{|c|}{Response}\\
\cline{2-8}
& Rigel & P-Rigel & P-Rigel & Rigel & P-Rigel & BFS & Rigel \\
\hline
Flickr & 1.4 & 0.028 & 0.003 & 9.7 & 0.24  & 24.5s &  12.9$\mu$s\\
Orkut & 7.5 & 0.15 & 0.005 & 19.4 & 0.42  & 56.2s & 36.6$\mu$s \\
Livejournal & 4.8 & 0.096 & 0.008 & 32.2 & 0.66 & 65.2s  & 8.4$\mu$s\\
Renren & 136.2 & 2.7 & 0.07 & 250 & 6.4 & 1598.5s & 28.9$\mu$s\\
\hline
\end{tabular}
\caption{\small Comparing the time complexity of Rigel and Parallel Rigel
  (P-Rigel) using a cluster of 50 servers. 
 The parallelization reduces the embedding time by
  nearly a factor of 50.  Compared to BFS, the per-query response time of
  both Rigel and Parallel Rigel is at least 8 orders of magnitude lower.}
\label{tab:runningtime}
\end{table*}

\subsection{Parallelizing Graph Embedding}
Parallelizing Rigel is feasible because of two reasons. First, landmark
bootstrapping requires computing BFS trees rooted from each landmark, which
can be run independently and in parallel on different servers. Second, after
bootstrapping, each graph node $u$ can also be embedded independently and in
parallel based on the coordinates of the global landmarks. Because the number
of nodes is large, we just need to distribute nodes across servers to ensure
load balancing.

\begin{figure}[t]
\centering
   \epsfig{file=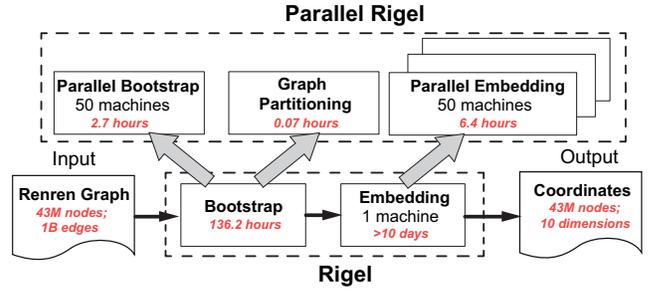,width=3.3in}
\vspace{-0.1in}
   \caption{A high-level view of how embedding is parallelized and its
     net impact on embedding latency for Renren, our largest graph.}
   \label{fig:wf}
\end{figure}

\para{Parallel Rigel.}  We integrate the above mechanisms with the original
Rigel design. The result is {\em Parallel Rigel}, an optimized version of
Rigel that scales to massive social graphs.  Figure~\ref{fig:wf} demonstrates
the Parallel Rigel system on top of and contrasts it to the original Rigel
design. It consists of three components: {\em parallel bootstrapping}, {\em
  graph partitioning} and {\em parallel embedding}. The parallel
bootstrapping module distributes BFS tree computation related to each
landmark across servers, one or more landmarks per server.  The graph
partitioning module provides a balanced distribution of nodes across
servers. The cost of this operation is negligible since simple partitioning
schemes are sufficient.  Finally, the parallel embedding module
embeds all graph nodes in parallel across the servers, allowing Parallel
Rigel to achieve significant speedup.


We have implemented a fully-functional prototype of parallel Rigel, and used
it to embed the largest graph we have, the 43 million node graph from the
Renren online social network.  As seen in Figure~\ref{fig:wf}, running the
centralized version of Rigel on a single large memory server (Dell PowerEdge
server with 32GB of RAM) required 136 hours to perform initial bootstrapping
(computing BFS trees), and more than 10 days to do the actual node embedding
of all graph nodes.  Applying parallel Rigel to the same graph over a cluster
of 50 servers (Dell Xeon, 2GB) reduces the parallel bootstrap process to 2.7 hours, and
embedding to only 6.4 hours.

\subsection{Experimental Results}
Using Parallel Rigel, we can now embed multi-million node graphs in a
reasonable amount of time. In the following, we use four of today's massive
social graphs,  Flickr, Orkut,
Livejournal and Renren, to examine the accuracy and efficiency of Parallel
Rigel. The characteristics of these four graphs are listed in Table~\ref{tab:graphs}.

\begin{figure}[t]
\centering
\epsfig{file=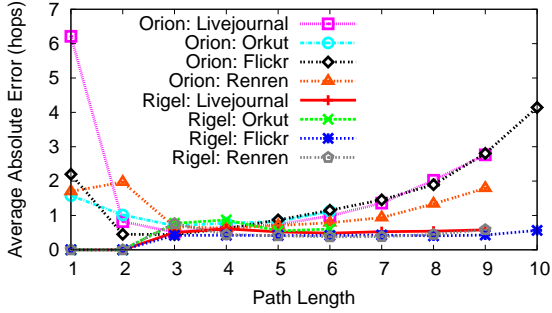, width=3in}
\caption{Average Absolute Error for different path lengths computed by
Parallel Rigel and Orion.}
\label{fig:largeab}
\end{figure}

\para{Accuracy.}  
We first examine the accuracy of Parallel Rigel's
coordinate system by comparing it to Orion.  In Figure~\ref{fig:largeab} we
plot the average absolute error for different path lengths using Parallel
Rigel and Orion. Like our previous results on smaller Facebook graphs,
Parallel Rigel not only significantly improves the accuracy of long distance
prediction, but also reduces the error in short distance estimation.  We have
also verified that Parallel Rigel performs similar to the original Rigel on
these graphs.

\para{Computation Efficiency.}  We now evaluate the efficiency of Parallel
Rigel by comparing its computation time against that of the original Rigel
design. By utilizing a cluster of servers, Parallel Rigel can distribute the
computation tasks of landmark bootstrapping and graph embedding into multiple
parallel servers. While Parallel Rigel does require an extra
step of graph partitioning by distributing nodes among machines, it only leads
to a minor increase in time complexity, less than 0.1\% of the original bootstrapping
time. Table~\ref{tab:runningtime}
shows the comparison when Parallel Rigel runs on a cluster of 50 servers.  We
see that  Parallel Rigel achieves close to linear speedup, even slightly
better due to better memory isolation across multiple servers.

\begin{figure}[t]
\centering
   \epsfig{file=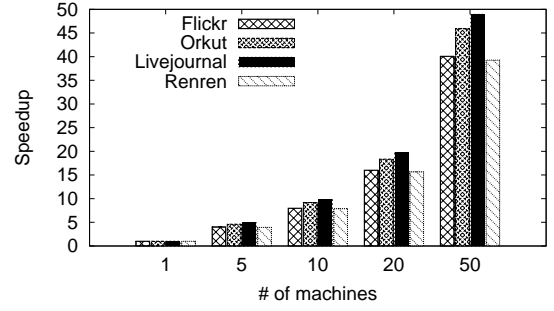,width=3in}
\caption {Average speedup achieved by Parallel Rigel on different
  cluster configurations.}
\label{fig:prtime}
\end{figure}

To examine the impact of the cluster size, we also compare the speedup
achieved by Parallel Rigel using 5, 10, 20 and 50 servers, where speedup is
the decrease in embedding time.  Results in Figure~\ref{fig:prtime} show that
run time decreases almost linearly with cluster size.

\section{Applications}
\label{sec:app}

We demonstrate the effectiveness and efficiency of Rigel in social network
analysis and applications by implementing several common graph applications.
In each case, we compare the accuracy of Rigel against that of
Orion~\cite{orion10}. 


\subsection{Computing Separation Metrics}
\label{sec:sep}
Social network graphs are known for displaying the ``Small World''
behavior. Graph separation metrics such as diameter, radius and average path
length, have been widely used to examine and quantify the Small World
behavior.  But since each of these metrics relies on large numbers of node
distance computations, computing them for large graphs can become extremely
costly or even intractable.



Using Rigel, we build an application to compute the graph separation metrics
listed above, and examine their accuracy by comparing their results to
ground truth.  Since computing shortest path length between all
node pairs takes several days even for our smallest graph (Facebook Egypt), we
take a random sampling approach to compute the ground truth.  We
randomly sample 5000 nodes from the three Facebook graphs, 500 nodes
from Flickr, Livejournal and Orkut, and $100$ nodes from Renren, and
use shortest path lengths between these pairs to derive the separation
metrics.

We report the results in Table~\ref{tab:sep} for Radius, Diameter and Average
Path Length on seven different graphs, for Rigel, Orion and Ground Truth.
In general, Rigel consistently provides more accurate results
compared to Orion.  More importantly, Rigel provides results across all
three metrics that are extremely close to ground truth values.

\begin{table*}[t]
\centering
\begin{tabular}{|c|c|c c c | c c c c |}
\hline
Metric & Method & Egypt & L. A. & Norway & Flickr & Orkut & Livejournal & Renren\\
\hline
\hline
\multirow{2}{13mm} {Radius} & Ground Truth & {\bf 9} & {\bf 11} & {\bf 8} &  {\bf 13} & {\bf 6} & {\bf 13} & {\bf 12} \\
\cline{2-9}
& Rigel & 8.7 & 11.0 & 7.5 &  12.7 & 6.4 & 12.2 & 12.0  \\
\cline{2-9}
& Orion  & 9.2 & 10.7 & 7.8 &  12.6 & 6.3 & 12.0 & 12.1\\
\hline
\hline
\multirow{2}{13mm}{Diameter} & Ground Truth &  {\bf 14} & {\bf 18} & {\bf 12} & {\bf 19} & {\bf 8} & {\bf 17} & {\bf 15}  \\
\cline{2-9}
&Rigel & 14.8 & 17.9 & 11.7 &  18.6 & 10.2 & 17.7 & 14.9 \\ 
\cline{2-9}
& Orion & 14.4 & 17.8 & 12.2 &  17.3 & 10.0 & 16.8 & 14.9 \\
\hline
\hline
\multirow{2}{13mm}{Average Path\\Length}& Ground Truth & {\bf 5.0} & {\bf 5.2} & {\bf 4.2} & {\bf 5.1} & {\bf 4.1} & {\bf 5.4} & {\bf 5.0}\\
\cline{2-9}
&Rigel & 4.9 & 5.1 & 4.2 & 5.0  & 4.3 & 5.5 & 4.9\\
\cline{2-9}
& Orion  & 4.7 & 5.0 & 4.1 & 4.3 & 3.9 & 4.8 & 4.6\\
\hline
\end{tabular}
\caption {Comparing separation metric results, as computed by Rigel,
  Orion, and BFS (ground truth).}
\label{tab:sep}
\end{table*}

\begin{figure*}[t]
\centering
\subfigure[L.A.]{
\epsfig{figure=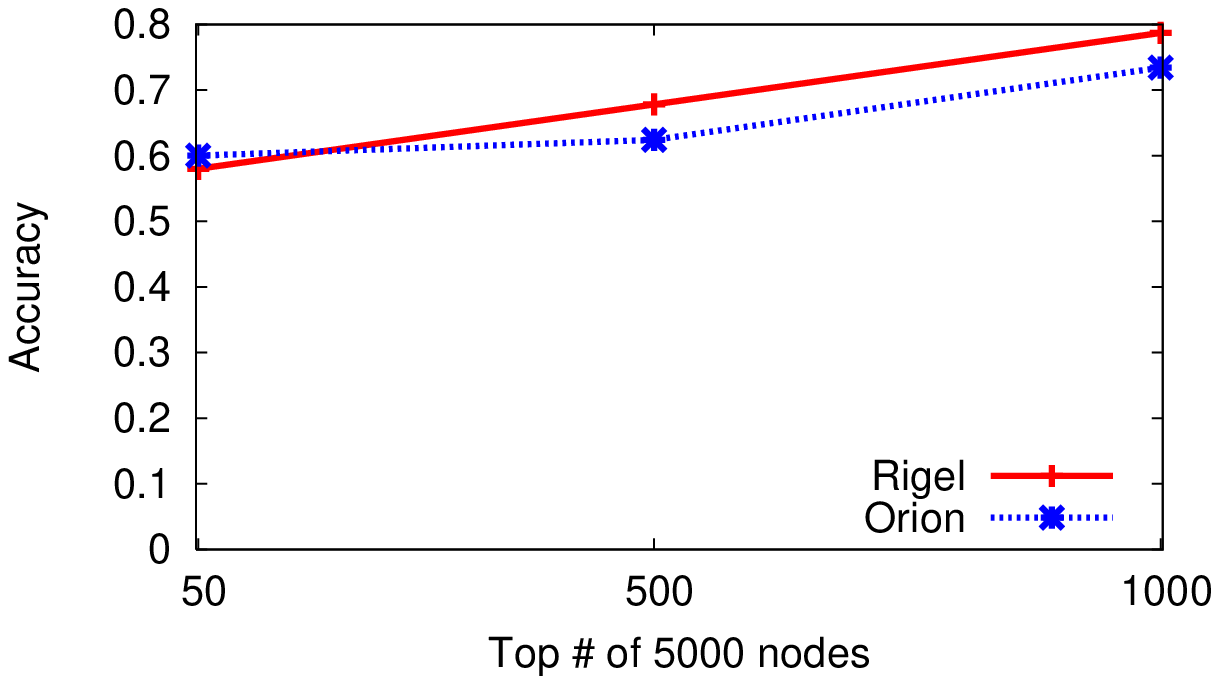, width=2.2in}
\label{fig:lacen}
}
\subfigure[Orkut]{
\epsfig{figure=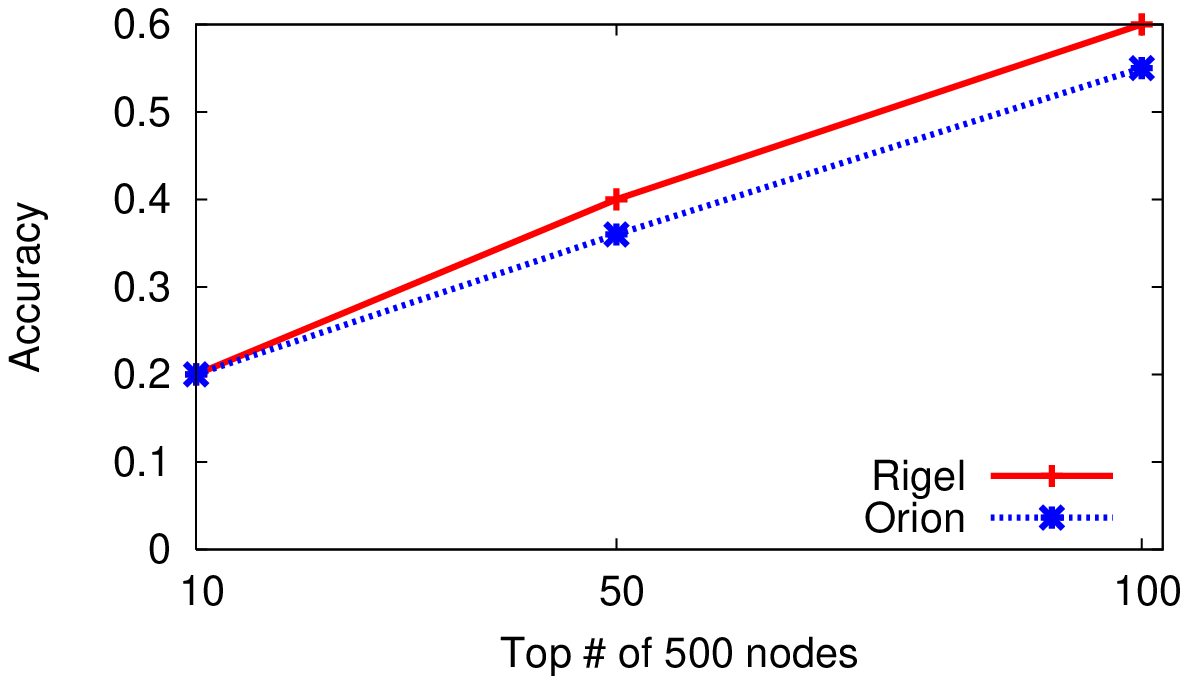, width=2.2in}
\label{fig:orkcen}
}
\subfigure[Livejournal]{
\epsfig{figure=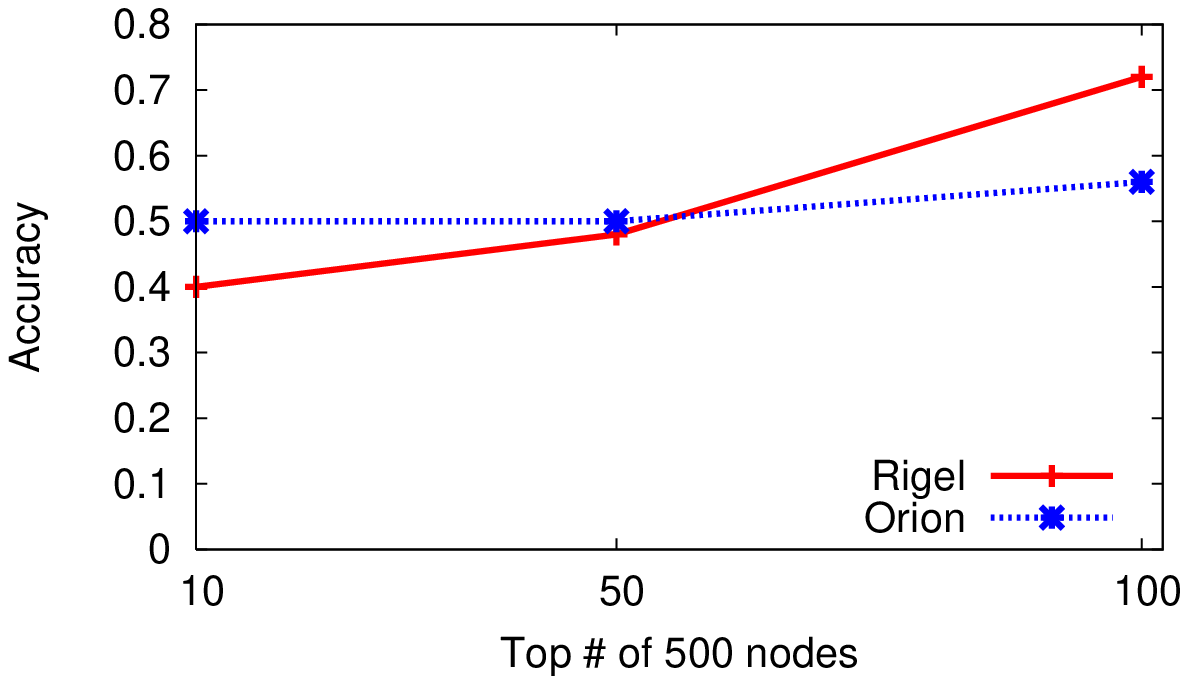, width=2.2in}
\label{fig:livecen}
}
\vspace{-0.1in}
\caption{Average accuracy of queries for the top $k$ high centrality
  nodes. Rigel consistently outperforms Orion. }
\label{fig:lcen}
\end{figure*}

\subsection{Computing Graph Centrality}
Graph centrality is an extremely useful metric for social applications such
as influence maximization~\cite{fast_max_influ} and social search. For
example, application developers can use node centrality values to identify
the most influential nodes for propagating information in an online social
network.  Formally, the most ``central'' node is defined as the node which has
the lowest average node distance to all other nodes in the network.  

Using Rigel, we implement a simple application to compute node centrality
directly from node distance values, where a node with a small average path
length has a high centrality score.  As before, we examine the accuracy of
our Rigel-enabled application by computing the centrality of $x=5000$ randomly
chosen nodes on the three Facebook graphs, $x=500$ randomly chosen nodes each for
Flickr, Livejournal Orkut, and $x=100$ nodes for Renren. 
For each graph, we sort these $x$
nodes by centrality, and select the top $k$ nodes.  We compute the
``accuracy'' of Rigel's centrality ordering by counting the number of
overlapping nodes ($m$) in Rigel's top $k$ nodes and actual top $k$
centrality nodes as computed by BFS on the original graph.  We study the
accuracy of our Rigel-based system as the ratio of $m$ to $k$.

We perform our experiments on all seven of our social graphs, and find the
results to be highly consistent.  For the rest of this section, we will only
report results for three of them: Facebook Los Angeles, Orkut and
Livejournal.  Figure~\ref{fig:lcen} shows the 
centrality accuracy results for different values of $k$.  As expected, the
accuracy of both Rigel and Orion increases with larger $k$ values.  In
general, Rigel consistently outperforms Orion for different graphs and
different values of $k$.

\begin{figure*}[t]
\centering
\subfigure[L.A.]{
\epsfig{figure=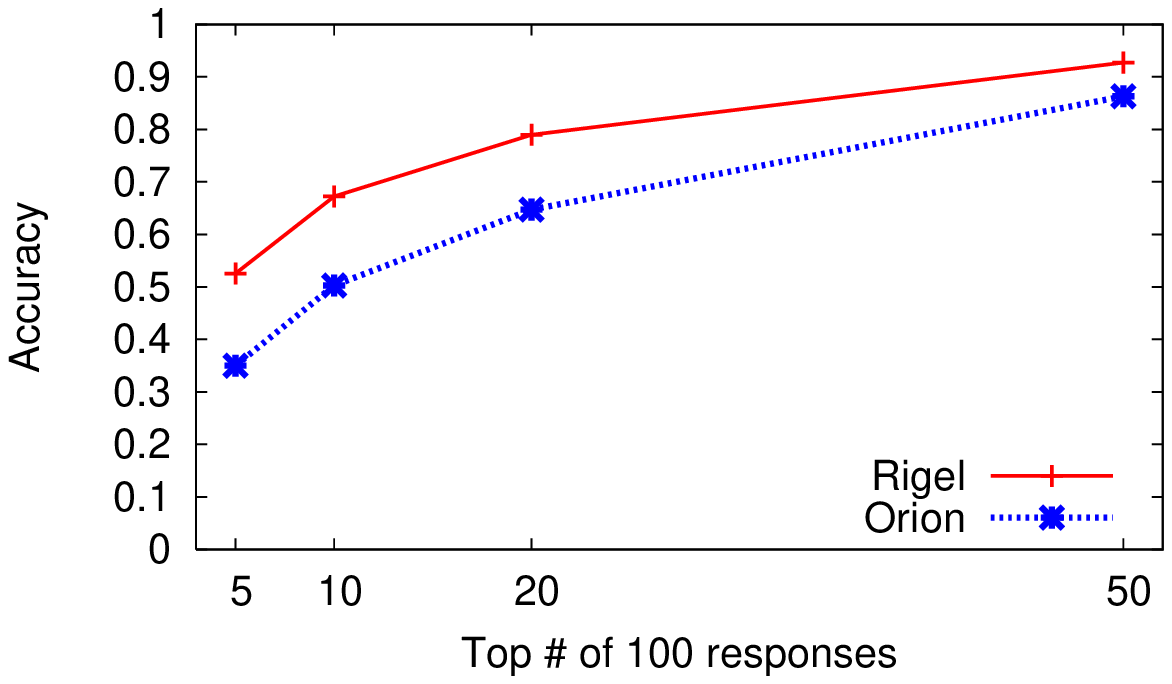, width=2.2in}
\label{fig:larank}
}
\subfigure[Orkut]{
\epsfig{figure=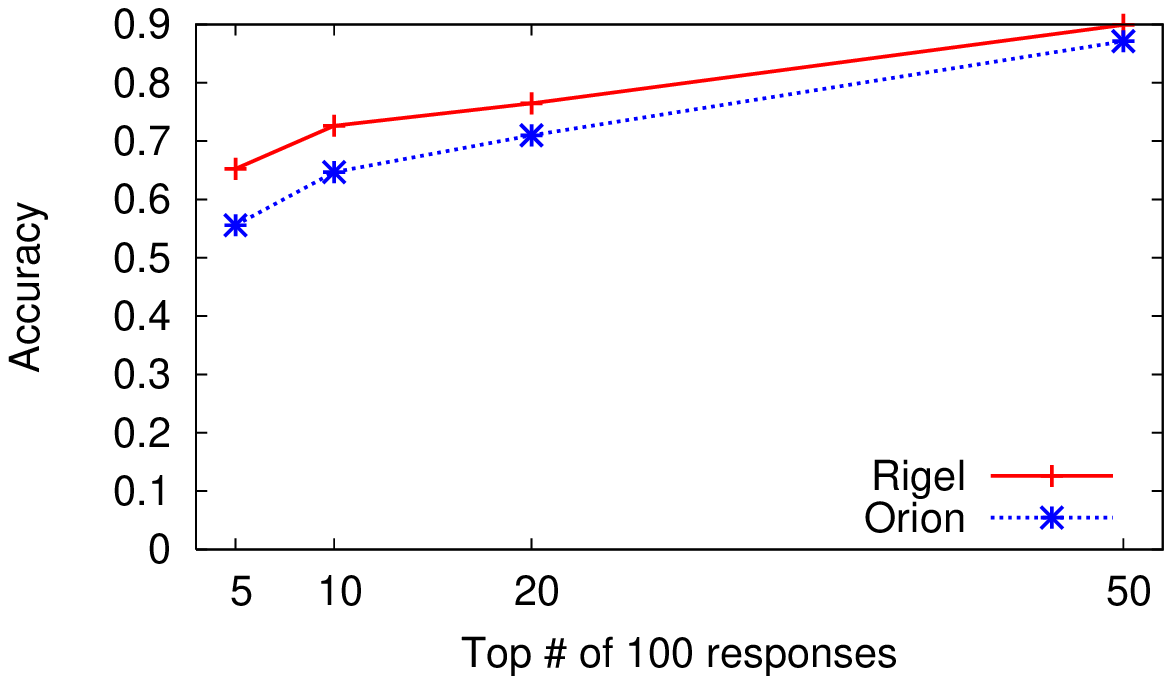, width=2.2in}
\label{fig:orkrank}
}
\subfigure[Livejournal]{
\epsfig{figure=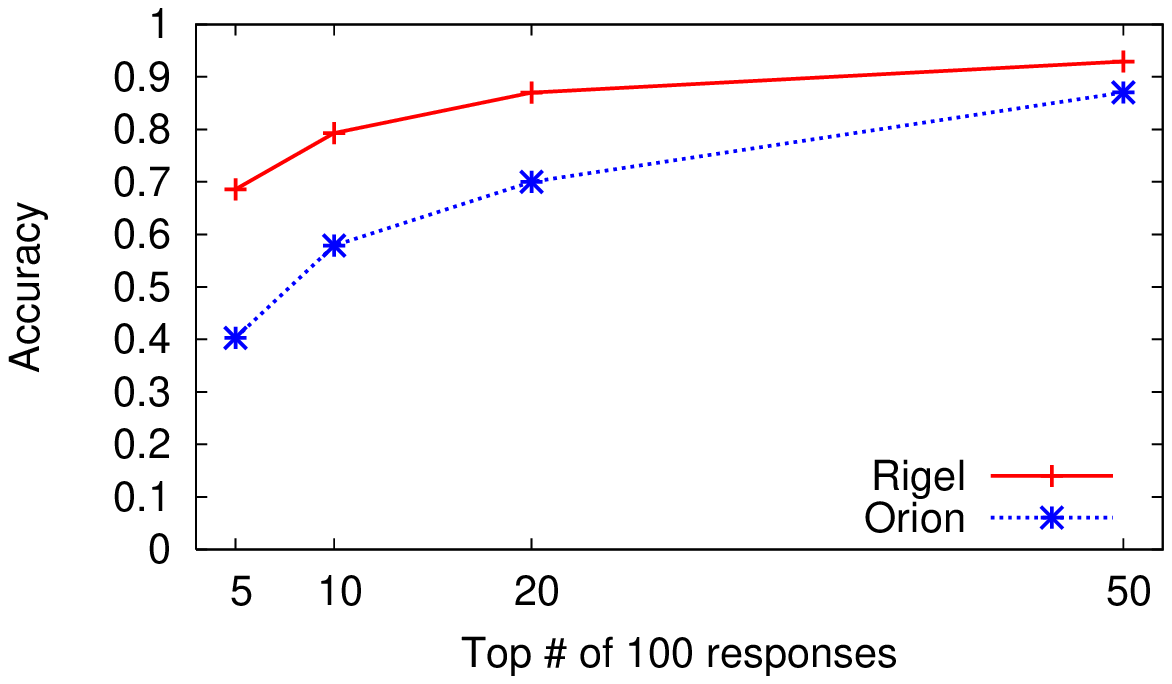, width=2.2in}
\label{fig:liverank}
}
\vspace{-0.1in}
\caption{Average accuracy of social search queries that return top $k$ ranked nodes}
\label{fig:lrank}
\end{figure*}

\subsection{Distance-Ranked Social Search}
Social networks such as Facebook and LinkedIn can best serve their users by
ranking search results by the proximity of each result to the user in the
social graph~\cite{socialsearch}. This is because users are likely to care about its
social proximity to the origin of the search result as much as the
quality of the result itself, {\em i.e.} a user would pay more interest
to results from her close friend rather than those from an unrelated
stranger.  

Despite its usefulness, including social distance in search results is highly
costly due to the number of node distance computations
necessary for each social
search query.  Instead, we can leverage Rigel's constant time node-distance
functionality to build powerful distance-based social search applications.

To verify the impact of Rigel on distance-ranked social search, we perform
the following experiment.  For each node which initiates a query, we randomly
select $100$ nodes in the network to respond to the query. We sort the
responses by their social distance to the query node, computed via both Rigel
and Orion, and return the top $k$ nodes for the user.  We then compute the
same top $k$ results by computing social distance using BFS, and examine the
percent of overlapping nodes between the result sets as a measure of
accuracy.  We repeat this experiment $5000$ times on smaller graphs like
Egypt, L.A. and Norway, and $100$ times on our largest graph, {\em i.e.}
Renren.  We vary the parameter $k$ from $5$ to $50$, and show the results of
L.A, Orkut and Livejournal in Figure~\ref{fig:lrank}.  The results show that
Rigel's hyperbolic coordinates allow it to consistently and significantly
outperform Orion's Euclidean coordinates. On Livejournal, for example,
when we rank the top $5\%$ search results, average accuracy of Rigel is $70\%$
while Orion only achieves $40\%$.

\section{Shortest Paths in Rigel} 
A number of critical graph-based applications require not only the length of
the shortest path between two nodes, but also the actual shortest path
connecting them.  For example, users often need to know the exact social
links that connect them to another user in LinkedIn.  Similarly, users in the
Overstock social auction system can perform a search to see how they are
connected to the seller of a given object~\cite{overstock}.

In this section, we describe a novel extension to Graph Coordinate Systems
that produces accurate approximations of shortest paths by using node
distance queries as a tool.  We first describe how this extension to Rigel
can compute short paths between any two nodes.  Next, we describe the Sketch
algorithm~\cite{sketch10}, an efficient algorithm for shortest path
estimation, and its followup algorithms including SketchCE, SketchCESC, and
TreeSketch~\cite{treesketch10}.  Finally, we compare Rigel's shortest path
algorithm against all of these algorithms on a variety of social graphs in
both accuracy and per-query runtime.  We show that while Rigel requires
similar preprocessing times to these algorithms, Rigel's shortest paths
return query results 3-18 times faster, while matching the best of these
algorithms in accuracy.

\subsection{Finding Shortest Paths using Rigel}
We now describe a heuristic that uses our coordinate system to find a good
approximation of the shortest path connecting any two nodes.  Our algorithm,
which we call {\em Rigel Paths}, uses techniques
reminiscent of the routing algorithm in~\cite{greedyforward10}.  

Given two nodes $A$ and $B$, we start by computing the distance between them
$d(A,B)$.  If the distance is 1 or 2 hops, we can use simple lookup on
their adjacency lists to determine the shortest path between them.  If the
predict distance between the nodes is greater than 2 hops, then we begin an
iterative process where we attempt to explore potential paths between the
nodes using the coordinate space as a directional guide.

Starting from $A$, we use Rigel to estimate the distance of each of its
neighbors $N^A_i$ to $B$.  The expected distance for a neighbor on the
shortest path should be $d(A,B)-1$.  If any neighbor's estimated distance is
within a $\delta$ factor of that prediction, it is considered a candidate
node to explore.  For each of $A$'s neighbors that qualify as a candidate
node, we repeat the process to obtain candidates for hop 2.  This process
iterates until one of the candidate nodes is a direct neighbor of $B$.

At each iteration of the algorithm, {\em i.e.} for the $n^{th}$ hop, we keep
a maximum number of candidates $C_{max}$ to explore.  Choosing this number manages the
tradeoff between exploring too many paths (and extending processing latency)
and exploring too few paths (and finding a dead end or inefficient paths). In
practice we choose $C_{max}$ to be 30, and $\delta$ to be 0.3.


\begin{figure*}[t]
\centering
\subfigure[L.A.]{
\epsfig{figure=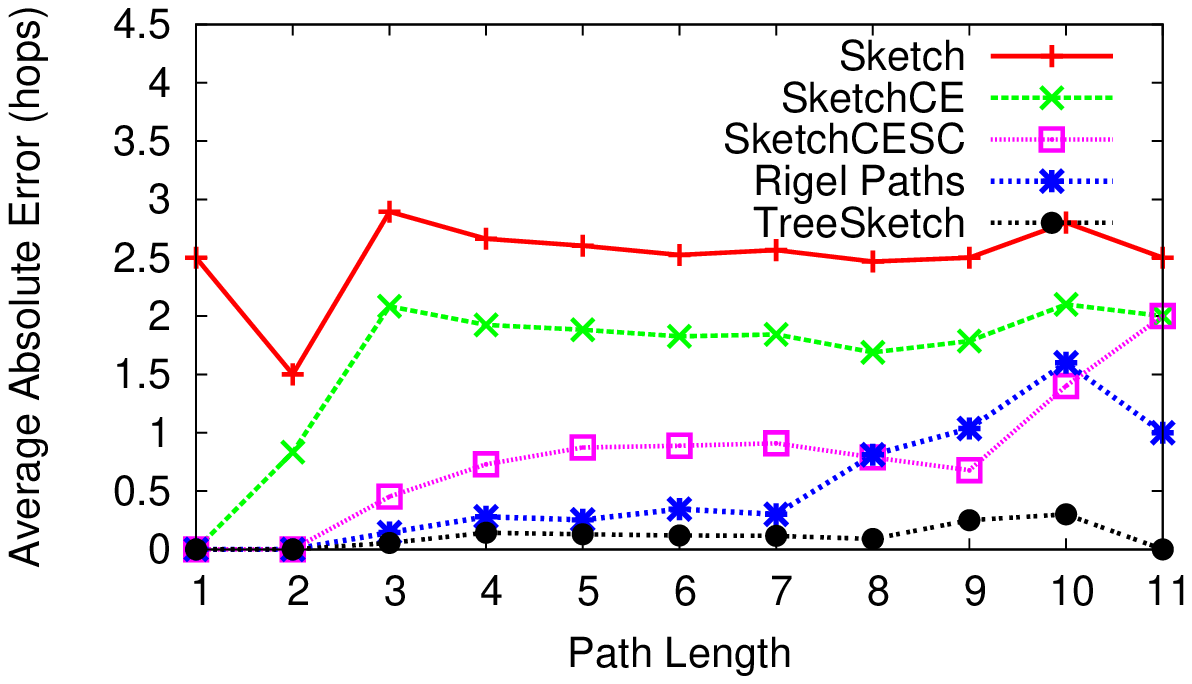, width=2.2in}
}
\subfigure[Orkut]{
\epsfig{figure=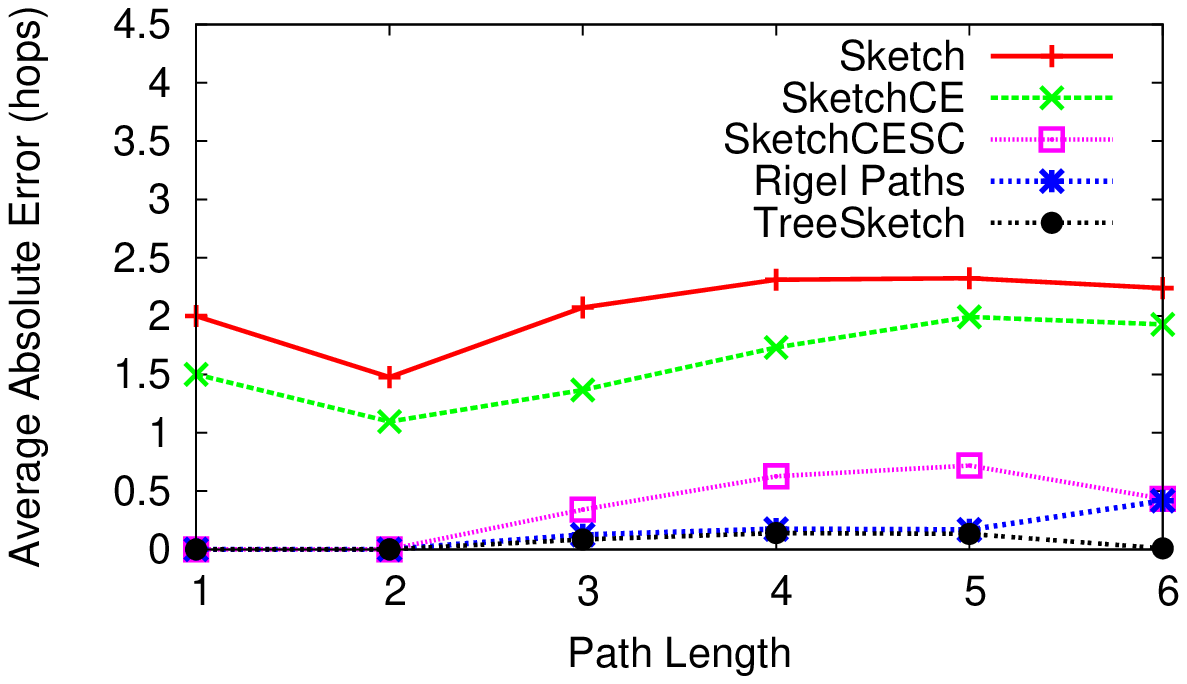, width=2.2in}
}
\subfigure[Livejournal]{
\epsfig{figure=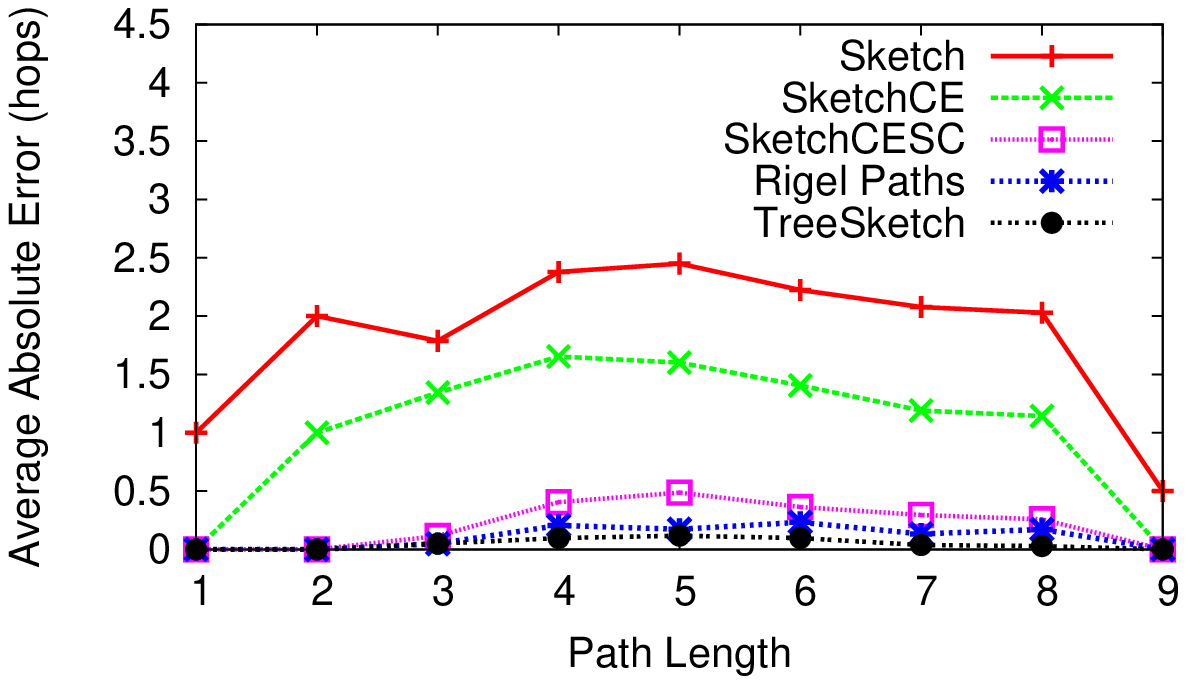, width=2.2in}
}
\vspace{-0.1in}
\caption{Absolute error (in hops) of shortest paths
  returned by Rigel Paths, Sketch, SketchCE, SketchCESC and TreeSketch.}
\label{fig:cikm}
\end{figure*}

\begin{figure*}[t]
\centering
\subfigure[L.A.]{
\epsfig{figure=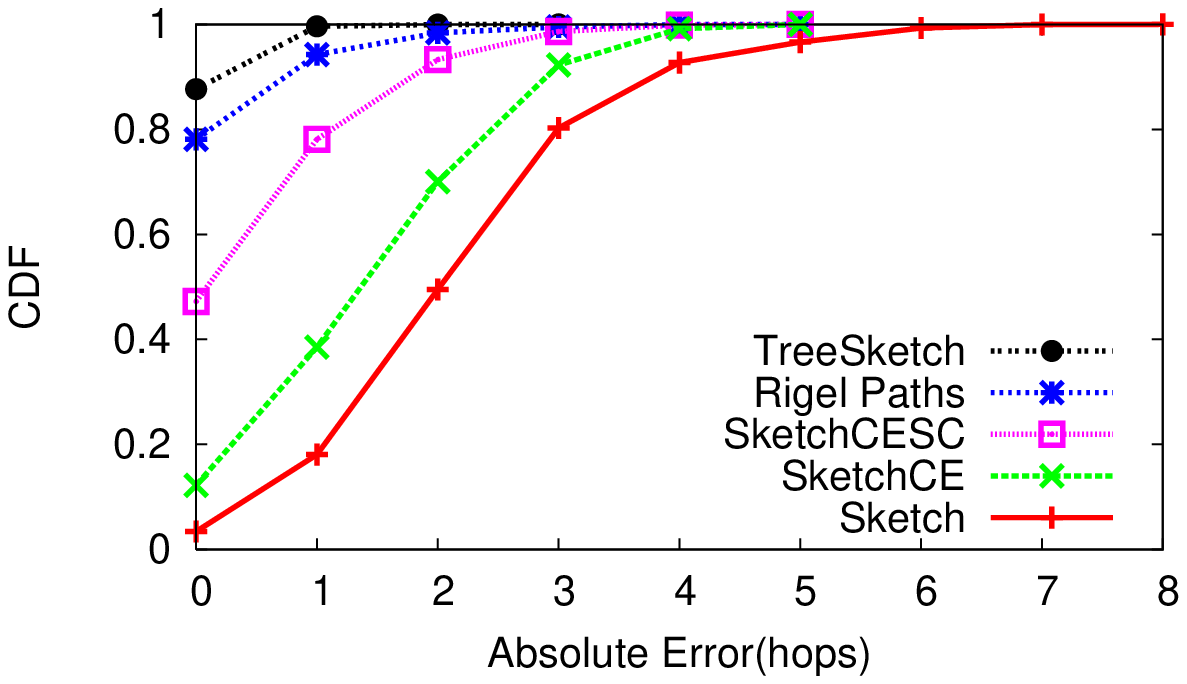, width=2.2in}
\label{fig:laerr}
}
\subfigure[Orkut]{
\epsfig{figure=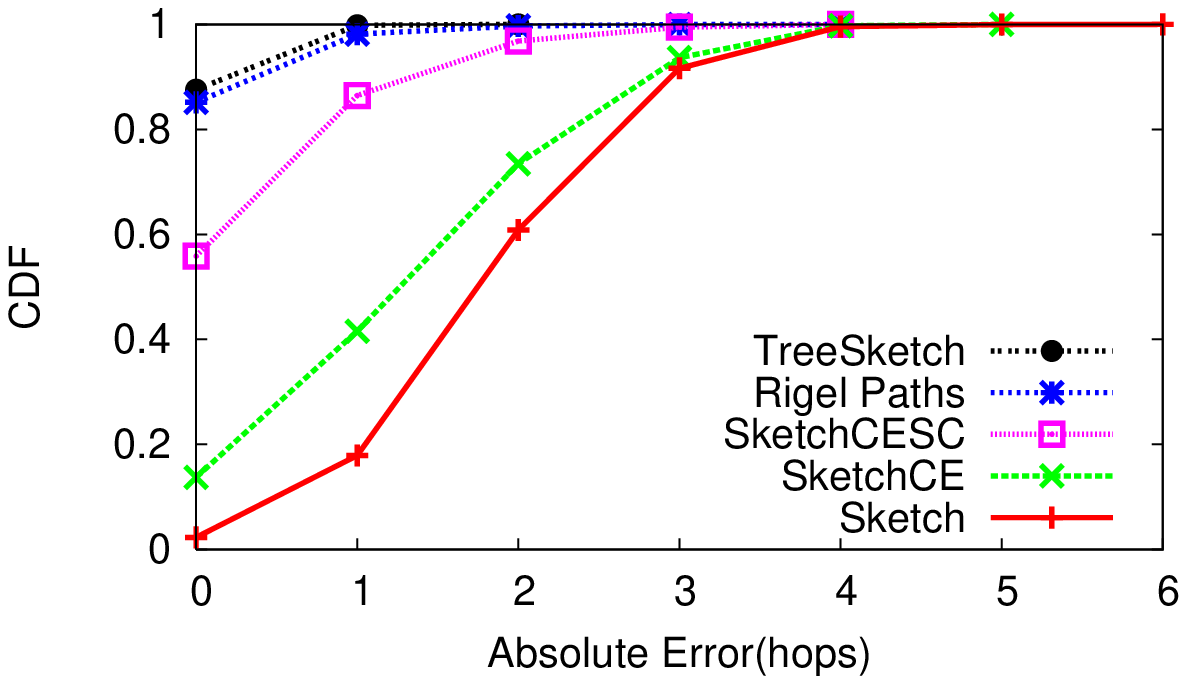, width=2.2in}
\label{fig:orkerr}
}
\subfigure[Livejournal]{
\epsfig{figure=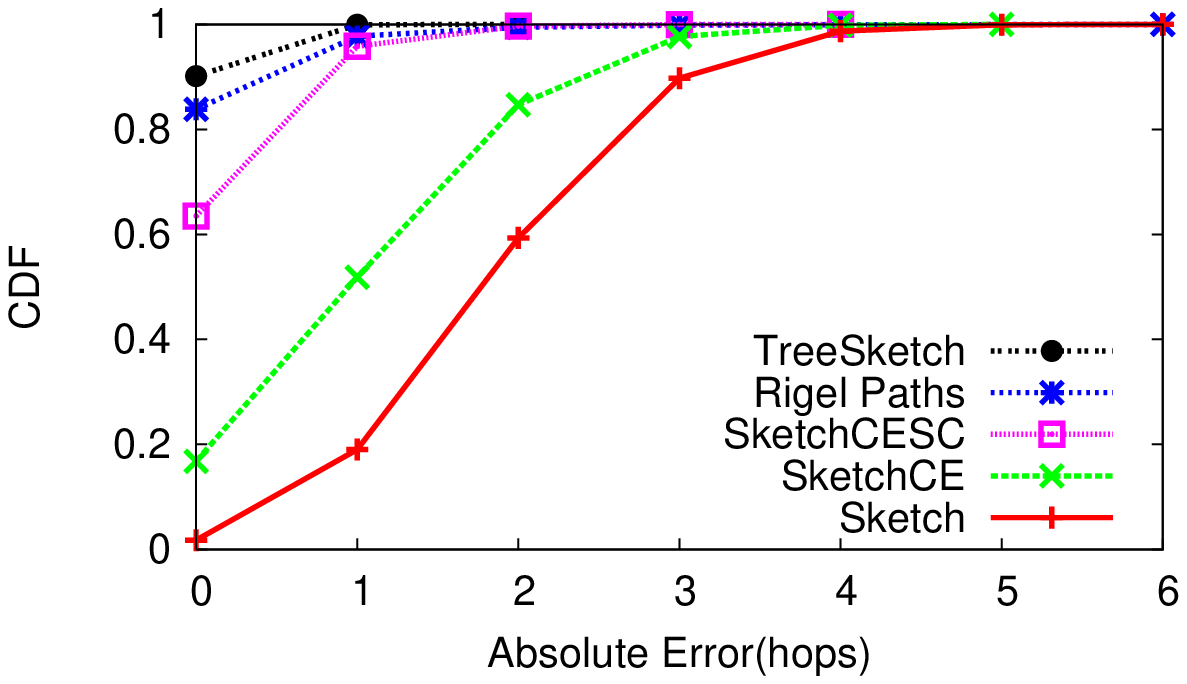, width=2.2in}
\label{fig:liveerr}
}
\vspace{-0.1in}
\caption{CDF of the absolute error in path finding among Rigel Paths,
    Sketch, SketchCE, SketchCESC and TreeSketch.}
\label{fig:patherr}
\end{figure*}


\begin{table*}[t]
\centering
\begin{tabular}{| c| c c | c c c c c c|}
\hline
Graphs & \multicolumn{2}{|c|}{Preprocessing (Hours)}  
& \multicolumn{6}{|c|}{Per-Query Response Time ($\mu$s)}\\
\cline{2-9}
& Rigel & Sketch & Rigel & Sketch & SketchCE & Rigel Paths & SketchCESC & TreeSketch \\
\hline
Egypt & 1.3  & 0.43   & 6.8 & 1781  & 1792 & 3667 &  38044 &
62407 \\
L.A. & 1.5 & 0.54  & 8.4 & 936  & 946 & 4008 & 20597 &
56828 \\
Norway & 1.4 & 0.67 &17.8 & 1492 & 1501  & 4621 & 21472
 & 59635\\
Flickr & 9.7 & 3.3  & 12.9 & 17157 & 17178  & 41279  & 732332  & 630890 \\
Orkut & 19.4 & 13.1   & 36.6 & 21043  & 21054& 49470 & 273586
 & 730284\\
Livejournal & 32.2 & 14.2  & 8.4 &  75101  & 75114 & 28355 & 253976
 & 348464 \\
Renren & 250 & 348 & 28.9 & 124327 & 124334 & 181814 & 546925 & 2594756\\
\hline
\end{tabular}
\caption{Comparing the preprocessing times and per-query response times of
  Rigel Paths, Sketch and variants SketchCE, SketchCESC and TreeSketch.
  Preprocessing/embedding time for Rigel (and Rigel Paths) is for single
  server (non-parallel version). Compared to the Sketch algorithms, 
  Rigel Paths reduces the per-query latency by a factor of 3
(against SketchCESC on Renren) to a factor of 18 (against SketchCESC on Flickr).}
\label{tab:cikmtime}
\end{table*}

\begin{figure*}[t]
\centering
\subfigure[L.A.]{
\epsfig{figure=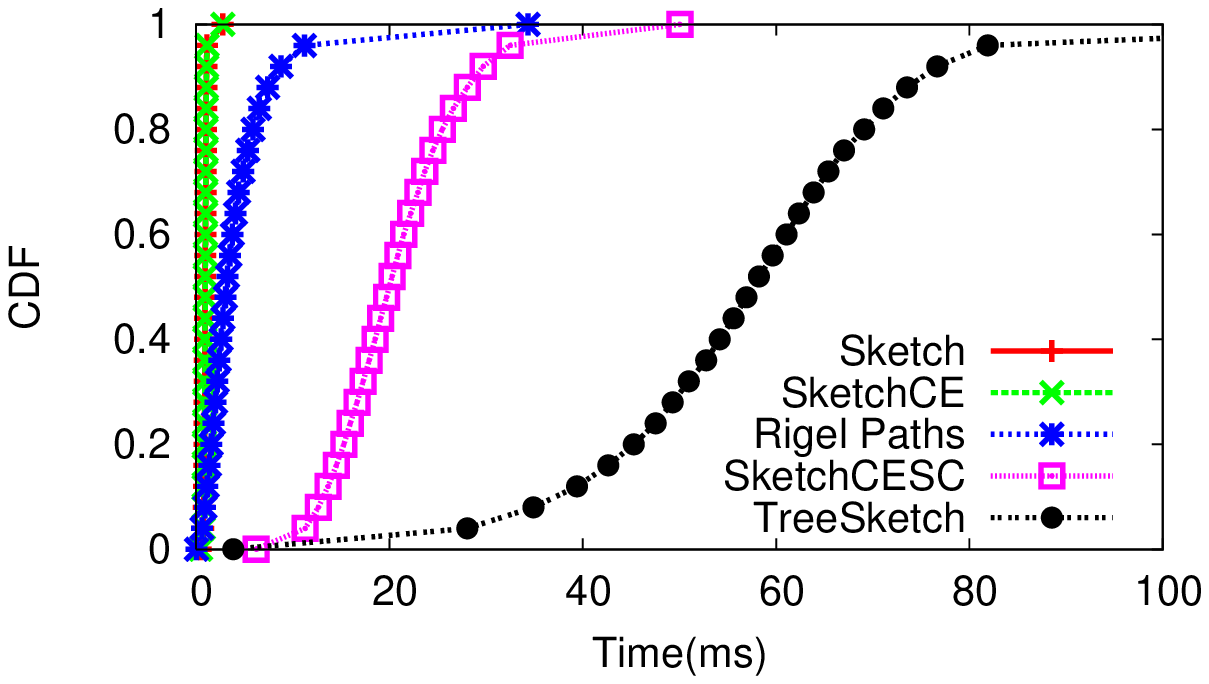, width=2.2in}
\label{fig:latime}
}
\subfigure[Orkut]{
\epsfig{figure=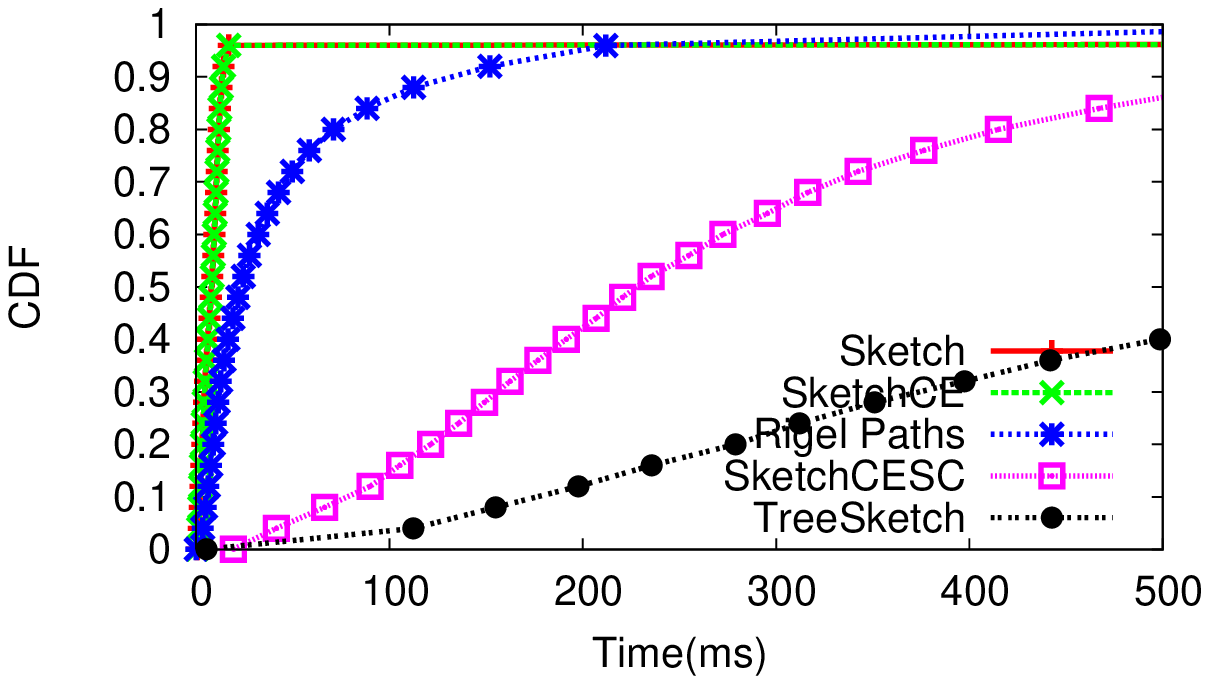, width=2.2in}
\label{fig:orktime}
}
\subfigure[Livejournal]{
\epsfig{figure=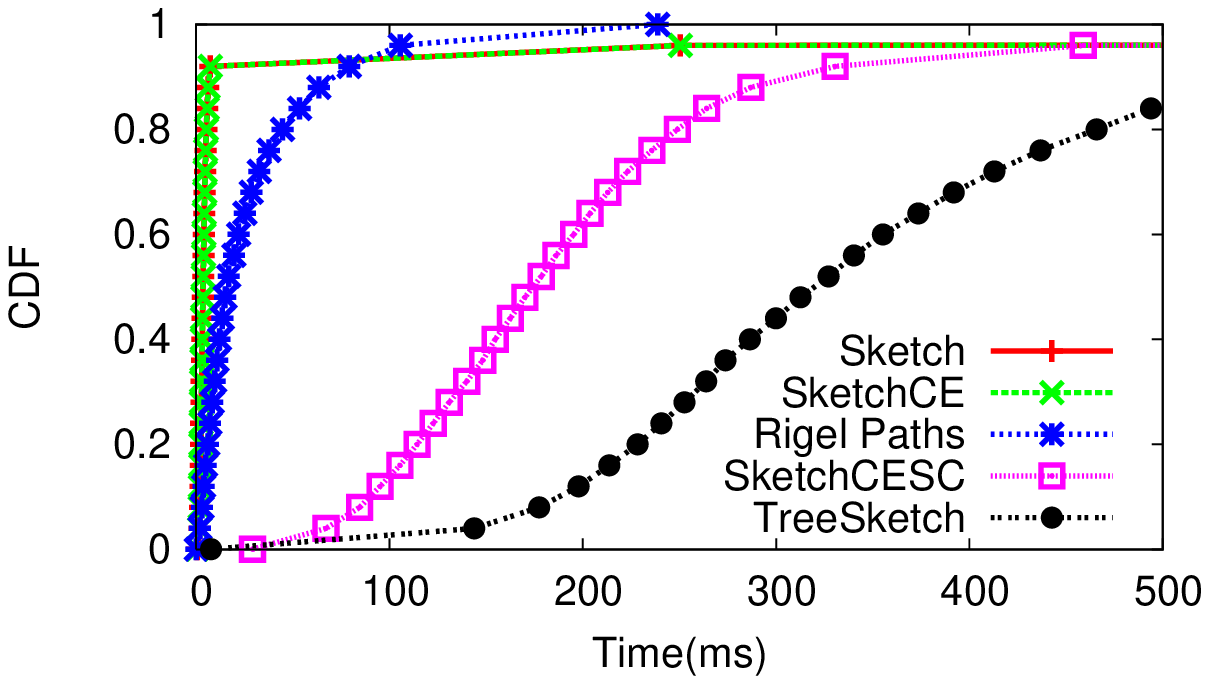, width=2.2in}
\label{fig:livetime}
}
\vspace{-0.1in}
\caption{CDF of computing time in path finding among Rigel Paths, Sketch,
    SketchCE, SketchCESC and TreeSketch.}
\label{fig:pathtime}
\end{figure*}

\subsection{Sketch-based Algorithms for Shortest Path}
\label{sec:Sketch_based_sortest_path}
Here, we describe existing state-of-the-art algorithms in approximating
shortest paths in graphs.  Two recent projects explored four total algorithms
for locating shortest paths, all based on variants of the Sketch
algorithm~\cite{sketch10,treesketch10}.  Here we describe these algorithms so
that we can compare them against Rigel in both accuracy and query latency.


\para{Sketch~\cite{sketch10}.} Sketch is a landmark-based solution where 
each node computes its shortest paths to the landmarks and then uses common
landmarks between itself and another node in the graph to estimate their
shortest paths.  This method selects $r=\lfloor{\log{ N }}\rfloor$ sets of
 landmark nodes, where $N$ is the number of nodes in the
graph.  For each node in the graph, Sketch computes its shortest paths to $k$~($k$=2)
different landmarks in each set~\cite{sketch10}.  Those shortest paths are precomputed by
leveraging the results of BFS trees rooted in each landmark.  Therefore,
for an undirected graph, each node is associated with $k\cdot r$ shortest paths.


\para{Cycle Elimination, Short Cutting and TreeSketch~\cite{treesketch10}.}
These three algorithms are variants of the basic Sketch approach for finding
shortest paths, and all three are described in \cite{treesketch10}.  {\em
  First,} Cycle Elimination improves Sketch by simply removing cycles in the
estimated path computed by Sketch.  We refer to this algorithm as SketchCE.
{\em Second,} Short Cutting improves Sketch by searching for bridging edges
between two nodes $x$ and $y$, where $x$ is on the path between the source
and the landmark and $y$ is on the path between the landmark and the
destination.  As soon as such an edge is found, the edge between $x$ and $y$
replaces the sub-path through the landmark. This approach also includes the
SketchCE optimization. It locates shorter paths, but dramatically increases
computational time.  We will refer to this algorithm as SketchCESC.

{\em Finally,} TreeSketch is a tree-based approach that improves Sketch by
adding another optimization to those implemented in SketchCE and SketchCESC.
At query time, TreeSketch builds two trees, one rooted at the source and one
rooted in the destination.  These trees are formed using precomputed
paths to landmarks; therefore, the computational time is proportional to the
complexity of building the trees and not to the BFS operations.
Given the two trees, the path search starts from both root nodes, and
iteratively explores more nodes from both trees.  BFS computation starts from
roots of both trees. For each visited node $u$ in a tree, its neighbors are
computed and compared with any visited node $v$ in the other tree. As soon as
a common node is found, the shortest path between source and destination is
constructed with the following three sub-paths: the subpath from source to
node $u$, the edge $(u,v)$, and the sub-path from $v$ to the destination.
While TreeSketch produces very accurate paths, it is computationally slow due
to the tree construction and extensive search process.


\subsection{Comparing Shortest Path Algorithms} 
\label{sec:camparing_accuracy/performance}
We compare our {\em Rigel Paths} algorithm to Sketch and its improved variants
SketchCE, SketchCESC and TreeSketch.  We compare both accuracy and per-query
latency. 

\para{Experimental Settings.} 
To compare Rigel Paths against prior work, we obtained the source code for the four
sketch-base algorithms from their authors~\cite{treesketch10}.  All of their
code runs on RDF-3X~\cite{rdf3x}, a specialized database system optimized for
efficient storage and computation of large graphs.  All graph experiments
were performed on Dell quad-core Xeon servers with 24GB of RAM, except for
Renren experiments, which were performed on similarly configured Dell servers
with 32GB of RAM.

\para{Accuracy.}
For each of the seven graphs in Table~\ref{tab:graphs}, we randomly sample 5000 node
pairs, and compare the shortest path results of Rigel Paths, Sketch, SketchCE,
SKetchCESC, and TreeSketch algorithms against the actual shortest paths
computed via BFS.  We evaluate the accuracy of these algorithms in two
ways.  First, we break down the 
absolute errors by the length of
the shortest path.  Second, we compute the estimate shortest paths, of the
5000 pairs of nodes, hop by hop and observe the similarity compared with the
ground truth.

Figure~\ref{fig:cikm} shows the average absolute error of the five different
algorithms broken down by length of the actual shortest path.  Here we define
the absolute error as the additional number of hops in the estimated path
when compared to the shortest path.  As before, we only show the Los Angeles
Facebook, Orkut and Livejournal graphs for brevity, because their results are
representative of results on other graphs.  The results show consistent
trends across the graphs.  The Sketch and SketchCE algorithms are highly
inaccurate, and generally produce shortest paths that are roughly 2 hops
longer than the shortest path.  TreeSketch and Rigel are the most accurate
algorithms.  They produce extremely accurate results, and are often
indistinguishable from each other. Both produce much more accurate
results than SketchCESC.



We show the CDF of absolute errors of the different algorithms in
Figure~\ref{fig:patherr}.  This shows a clearer picture of the
distribution of errors.  Again, Rigel paths and TreeSketch are by far the
most accurate algorithms.  Both produce exact shortest paths for a large
majority of node pairs.  Both are significantly better than SketchCESC.
SketchCE and Sketch are fairly inaccurate, and provide paths with multiple
hop errors for the overwhelming majority of node pairs.  While Rigel Paths provides
accuracy that matches or beats all of the Sketch based algorithms, we will show later
that it is significantly faster than both SketchCESC and TreeSketch (ranging
from a factor of 3 to a factor of 18 depending on the specific graph).

Finally, we also compared the length of the shortest paths found by our Rigel
Paths algorithm to node distance values estimated by Rigel. Interestingly,
Rigel Paths is more accurate, with absolute errors below 0.3, compared to
errors between 0.5 and 1 hop\footnote{We can observe this result by comparing
  Figure~\ref{fig:cikm} and Figure~\ref{fig:largeab}.}. 
Rigel Paths achieves this higher level of accuracy because it leverages
actual graph structure to compute its shortest paths.

\para{Computational Costs.}
A scalable system for analyzing large graphs requires both accuracy and
efficiency.  We now compare Rigel Paths and the four Sketch algorithms on computational
time complexity.  We break down our analysis into two components.  First, we
measure each algorithm's {\em preprocessing time}.  For Rigel Paths (and Rigel),
this represents 
the time required to embed the graph into the coordinate space, {\em i.e.}
computing coordinates for all nodes.  All Sketch algorithms share the same
bootstrapping process, which includes computing shortest paths (using BFS) to
all of their landmark nodes~\cite{treesketch10}.  Our second component
measures the computational latency required to resolve each query.  All
experiments are run on a single server.  As before, Renren experiments were
run on our 32GB RAM server, while all other experiments were run on identical
24GB RAM servers.

We summarize all of our timing results in Table~\ref{tab:cikmtime}.  Looking
at the bootstrap times, we see that Rigel  takes roughly 2--3 times longer to
preprocess.  Note, however, that these measurements only capture bootstrap
times for a single server.  As shown in Figure~\ref{fig:prtime}, we can
distribute Rigel's preprocessing phase across multiple machines with close to
linear speedup.  Once we consider this factor, we see that we can reduce
Rigel preprocessing to match Sketch just by spreading the load over 2 or 3
machines. 

Per-query latency is likely to be a much more important measure of
performance, since large social graphs are unlikely to change significantly
over short time periods.  Again, we choose 5000 node pairs at random from
each of the graphs, and compare the average query response time for each of
the algorithms.  The shortest path algorithms, Sketch, SketchCE, Rigel Paths,
SketchCESC and TreeSketch are ordered in Table~\ref{tab:cikmtime} from left
to right from the fastest to the slowest.  Recall from prior results that
Sketch and SketchCE produce paths that are highly inaccurate, {\em i.e.}
introduce an average of 2-3 additional hops in each path.  Of the two best
algorithms, Rigel Paths and TreeSketch, Rigel paths returns results in a
fraction of the time required by TreeSketch and SketchCESC.  The latency
reduction ranges from $\sim$3 (against SketchCESC on Renren) to a factor of
18 (against SketchCESC on Flickr).  We show a CDF of these results in
Figure~\ref{fig:pathtime}. Rigel Paths is clearly much faster than both
TreeSketch and SketchCESC.

Finally, we also include the node-distance computation time from Rigel as
a point of reference.  Clearly, finding actual shortest paths is orders of
magnitude more expensive than simply computing node distance. Luckily, the
large majority of graph analysis tasks only require node-distance
computation, and only user-interactive queries require the full shortest path
between node pairs.



\vspace{-0.1in}
\section{Conclusion}
\label{sec:con}

Traditional algorithms for performing graph analytics no longer scale to
today's massive graphs with millions of nodes and billions of edges.
Computing distances and shortest paths between nodes lies at the heart
of most graph analysis metrics and applications, and is often responsible for
making them intractable on large graphs. 

We propose Rigel, a hyperbolic graph coordinate system that approximates node
distances by first embedding graphs into a hyperbolic space.  Even for graphs
with 43 million nodes and 1+ billion edges, Rigel not only produces
significantly more accurate results than prior system, but also answers node
distance queries in 10's of microseconds using commodity computing servers.
For the more challenging task of computing shortest paths, we propose Rigel
Paths, a highly efficient algorithm that leverages Rigel's node distance
estimates to locate shortest paths.  The results are impressive.  Rigel Paths
produces exact shortest paths for the large majority of node pairs, matching
the most accurate of prior systems.  And it does this quickly, returning
results up to 18 times faster than state-of-the-art 
shortest-path systems with similar levels of accuracy.


\begin{small}
\bibliographystyle{IEEEtran}
\bibliography{zhao,social,p2p}
\end{small}

\end{document}